\documentclass[fleqn,usenatbib]{mnras}
\newcommand\HII{$\textrm{H}\scriptstyle\mathrm{II}$}

\usepackage[T1]{fontenc}
\usepackage[utf8]{inputenc}

\DeclareRobustCommand{\VAN}[3]{#2}
\let\VANthebibliography\thebibliography
\def\thebibliography{\DeclareRobustCommand{\VAN}[3]{##3}\VANthebibliography}

\usepackage{graphicx}	
\usepackage{amssymb}	
\usepackage{newtxtext,newtxmath}

\title[Early X-ray feedback]{Maximal X-ray feedback in the pre-reionization universe}

\author[Jeon J. et al.]{
Junehyoung Jeon\textsuperscript{\href{https://orcid.org/0000-0002-6038-5016}{\includegraphics[width=2.5mm]{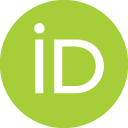}}\,}\thanks{E-mail: junehyoungjeon@utexas.edu}$^{1}$, 
Volker Bromm\textsuperscript{\href{https://orcid.org/0000-0003-0212-2979}{\includegraphics[width=2.5mm]{orcid.png}}}$^{1}$ 
and Steven L.~Finkelstein\textsuperscript{\href{https://orcid.org/0000-0001-8519-1130}{\includegraphics[width=2.5mm]{orcid.png}}}$^{1}$
\\
$^{1}$Department of Astronomy, University of Texas, Austin, TX 78712, USA\\
}

\date{Accepted 2022 July 28. Received 2022 July 27; in original form 2022 June 09}

\pubyear{2022}

\begin{document}
\label{firstpage}
\pagerange{\pageref{firstpage}--\pageref{lastpage}}

\maketitle{}

\begin{abstract}
X-ray feedback in the pre-reionization Universe provided one of the major energy sources for reionization and the thermal evolution of the early intergalactic medium. However, X-ray sources at high redshift have remained largely inaccessible to observations. One alternative approach to study the overall effects of X-ray feedback in the early Universe is a full cosmological simulation. Toward this goal, in this paper we create an analytic model of X-ray feedback from accretion onto supermassive black holes (SMBHs), to be used as a sub-grid model in future cosmological simulations. Our analytic model provides a relation between the mass of a dark matter halo and the SMBH it hosts, where the efficiency is governed by an energy balance argument between thermal feedback and the confining gravitational potential of the halo. To calibrate the model, we couple the halo-level recipe with the Press-Schechter halo mass function and derive global mass and energy densities. We then compare our model to various observational constraints, such as the resulting soft X-ray and IR cosmic radiation backgrounds, to test our choice of model parameters. We in particular derive model parameters that do not violate any constraints, while providing maximal X-ray feedback prior to reionization. In addition, we consider the contribution of SMBH X-ray sources to reionization and the global 21\,cm absorption signal.
\end{abstract}





\section{Introduction}\label{1}
In the early Universe, sources of X-ray radiation had a significant effect on cosmic history \citep[][]{Mesinger2013}. First, X-ray and UV radiation drove the early stages of reionization \citep{Glover2003,Baek2010}. Noticeable differences in the timing and topology of reionization occur depending on the leading radiation source. When UV sources dominate, reionization is thought to occur in bubbles of ionized gas within a neutral background \citep[e.g.,][]{Loeb2001,McQuinn2007}. However, when X-ray sources dominate, the ionized regions are distributed more evenly due to the large mean free path of X-rays \citep{Fragos2013, Pritchard2007}. Furthermore, the X-ray radiation would have delayed reionization by increasing the Jeans mass through preheating, so that the timescale for halo formation was longer \citep{Jeon2014,Finkelstein2019}. On the contrary, X-ray radiation could suppress small-scale structure formation, thus in turn reducing recombination losses of ionizing photons, to accelerate reionization \citep{Finlator2012,Emberson2013}. However, there is still considerable uncertainty on what sources dominate reionization. The general view is that UV sources dominate and drive reionization \citep[e.g.,][]{Ciardi2003,Parsa2018,Finkelstein2019,Robertson2015,Naidu2020}, while some studies stress the role of active galactic nuclei (AGN) in completing reionization \citep[e.g.,][]{Madau2015,Volonteri2016}. 

The overall impact of X-ray feedback on star formation in the early Universe is complex. On the one hand, sources heat nearby gas and thus hinder star formation \citep{Jeon2014}. However, at larger distances, X-ray feedback may be able to compensate for the negative feedback from the UV photo-dissociation of molecular hydrogen, the main coolant in primordial star forming haloes. The partial ionization from X-rays can result in an enhanced abundance of free electrons in predominantly neutral hydrogen gas, in turn catalyzing the formation of hydrogen molecules \citep{Glover2003}. The net effect can be additional cooling in protogalaxies, thus enhancing the early star formation rate. 

Moreover, from the observed AGN X-ray and bolometric luminosities, the black hole (BH) accretion rate can be inferred if a constant radiative efficiency is assumed. Previous work found evidence for a correlation between the BH accretion rate and the star formation rate (SFR) for bulge-dominated galaxies \citep{Yang2019}, possibly because the supermassive black holes (SMBHs) coevolve with the stellar component of host galaxies \citep[e.g.][]{Kormendy2013,Davis2018}. This correlation can be used to determine the BH accretion density, to constrain models of BH formation and evolution \citep{Yang2021}. However, the known accretion density inferred from X-ray observations is an order of magnitude lower than what is predicted in theoretical simulation results \citep[e.g.,][]{Crain2015, Weinberger2017,Volonteri2016}. The reason for the discrepancy may be the observational bias due to obscured AGNs at high redshift \citep{Weigel2015,Vito2016,Yang2021}. Therefore, the nature of X-ray feedback, its sources and overall strength, needs to be better understood to deduce the progress of reionization, the formation and evolution of protogalaxies, and the implications from observations with frontier facilities, such as the \textit{James Webb Space Telescope} (\textit{JWST}). 

Although X-ray feedback has been considered in many simulations of the early Universe \citep[e.g.,][]{Ahn2015, Fragos2013,Monsalve2019,Jeon2014}, its overall amplitude is highly uncertain, mainly due to the current lack of direct observations \citep{Reines2016}. Here, we therefore seek to determine the maximum X-ray feedback allowed by existing cosmological constraints, to be employed as a `maximal' sub-grid model in future cosmological simulations. Direct constraints are the SMBH mass density at lower redshifts and the quasar luminosity function. For the SMBH mass density, we will consider local measurements \citep[e.g.][]{Fukugita2004,Graham2007,Marconi2004}, as well as semi-empirical determinations around $z\sim4$ \citep[e.g.][]{Mahmood2005,Tucci2017}. While we model sources at $z\sim5-15$, beyond current observational capabilities, we compare extrapolations of our model to existing observations of the quasar luminosity function at $z\sim5$ \citep[e.g.][]{Shen2020}. Furthermore, X-ray sources have a significant impact on reionization, as stated previously, and they also indirectly affect the signal from 21\,cm cosmology. Finally, X-ray feedback has to be consistent with cosmic radiation backgrounds, including the soft X-ray background\footnote{We do not consider the hard X-ray background, as the AGN that produce this component peaked at $z\sim0-3$, outside the time period considered here \citep{Matteo1999,Mushotzky2000}.}. We further consider the cosmic infrared background (CIB) as an indirect constraint, as the CIB is a global repository of reprocessed ionizing radiation at high redshifts, including the ionization resulting from X-ray photons. 

Our resulting model of maximal X-ray feedback in the early Universe can be tested and improved with upcoming frontier facilities over a large range of wavelengths. Among them are the Euclid mission to study the CIB \citep{Kashlinsky2018}, the Square Kilometer array to map the cosmological 21\,cm line \citep{Ghara2017}, the \textit{JWST} to observe AGN emission from accretion at high redshifts \citep{Volonteri2017}, and the Athena and Lynx X-ray observatories \citep{Barret2013,Gaskin2019} to directly observe high redshift AGN and quasar X-ray sources.

This paper is organized as follows. In Section \ref{2}, we discuss our modeling of X-ray feedback. We first model the number density and energy output for each X-ray source. In Section \ref{3}, we discuss the cosmological constraints considered here: reionization, the quasar luminosity function, the soft X-ray background, the CIB, and the 21\,cm line. We conclude by summarizing our results in Section \ref{5}.

\section{Modeling X-ray Feedback}\label{2}
We will model X-ray feedback in the pre-reionization Universe by considering AGN/SMBH and X-ray binaries \citep{Mirabel2011,Baek2010,Jeon2014,Fragos2013}. For SMBH sources, we consider `standard' formation pathways from massive stellar seeds, and will only briefly comment on the alternative direct collapse black hole (DCBH) channel, as the conditions for DCBH seed formation may be rare and difficult to achieve \citep{Bromm2003,Habouzit2016}. 

\subsection{Number and Mass Density of Sources}
\label{sec:maths} 
\subsubsection{AGN/SMBH Sources}
To derive the AGN/SMBH number density, we employ the Press-Schechter (PS) formalism \citep{Press1974}. More specifically, the PS formalism provides the mass function of dark matter haloes, evaluated with the Colossus package for Python \citep{Diemer2018}. When integrating the PS function over mass to obtain total mass densities, we consider an upper halo limit of $10^{13}M_{\odot}$, following prior work \citep[e.g.][]{Shankar2006,Angles2017}. For simplicity, we assume that each halo contains a black hole with mass proportional to the halo mass, such that

\begin{equation}\label{e1}
    M_{\text{BH}} = \eta M_{H}\mbox{\ .}
\end{equation}

To determine $\eta$, we consider a global energy balance argument. The fraction of the BH rest mass energy that is converted into radiation and is subsequently absorbed to heat the gas should equal the potential energy of the gas confined in the gravitational potential well of the host halo:
\begin{equation}\label{e2}
    M_{\text{gas}}\frac{GM_H}{R}  \simeq fc^2M_{\text{BH}}\mbox{\ .}
\end{equation}
Here, $R$ is the virial radius of the halo, and $f$ the fraction of the BH rest energy that will be deposited locally into the halo gas as heat energy.
Assuming that $M_\text{gas}\simeq (\Omega_b/\Omega_m) M_H$, where $\Omega_b/\Omega_m\simeq 0.16$ is the cosmological baryon-to-mass ratio for density parameters in baryons and matter of $\Omega_b=0.049$ and $\Omega_m=0.31$, respectively \citep{Plank2016}, and applying the relation between virial mass and radius of the halo:
\begin{equation}\label{e3}
    M_H = \frac{4}{3}\pi(200\rho_c(z))R^3\mbox{\ ,}
\end{equation}
where $\rho_c(z)$ is the background density of the Universe at redshift $z$, we can express the BH efficiency as follows 

\begin{equation}\label{e4}
   \eta\simeq 2\times10^{-7}\left(\frac{1+z}{10}\right)\left(\frac{f}{0.1}\right)^{-1}\left(\frac{M_H}{10^{10} M_\odot}\right)^{2/3}\mbox{\ .}
\end{equation}

To complete our prescription, we need to further model the energy conversion factor, which is expected to be of order $f\sim 0.1$ for the lower range of AGN radiative efficiencies \citep{Davis2011}. In somewhat more detail, we assume that accretion onto the SMBH is proceeding at the Eddington rate, consistent with observations of high-$z$ quasars that they radiate at the Eddington luminosity \citep[e.g.][]{Mortlock2011,Wu2015,Banados2018}.
Assuming that the accretion disk is truncated at the radius of the last stable orbit for a non-rotating central BH, the resulting luminosity is
\begin{equation}\label{e7}
    L \simeq 0.1\dot{M}_\text{Edd}c^2\mbox{\ .}
\end{equation}
Further assuming a feedback-limited duty cycle of $t_\text{feed}$ over which the SMBH is active, the feedback energy deposited locally as heat can be rewritten as 
\begin{equation} \label{e8}
    0.1(1-e^{-\tau})\dot{M}_\text{Edd}c^2t_{\text{feed}} = fM_{\text{BH}}c^2
    \mbox{\ ,}
\end{equation}
where $\tau$ is the optical depth of the halo gas. Within the framework of Eddington-limited BH growth, we can equate the AGN duty cycle with the Salpeter time, $t_\text{feed}\sim t_\text{Sal}\simeq M_\text{BH}/\dot{M}_\text{Edd}$ \citep[e.g.][]{Woods2019}. The optical depth to absorption can be estimated as  $\tau = R/\lambda_{\text{mfp}}$, where $\lambda_{\text{mfp}}$ is the mean free path for a photon emitted by the AGN traveling through the surrounding halo gas. The latter has an approximate number density of $0.16\times200\rho_c(z)/m_\text{H}$, where $m_\text{H}$ is the hydrogen mass. Using a radiative efficiency of 0.1, the feedback efficiency and optical depth become
\begin{equation}\label{e9}
    f \simeq 0.1(1-e^{-\tau})
\end{equation}
\[
\tau =  \left(\frac{3M_H}{800\pi}\right)^{1/3}\frac{0.16\times200\sigma}{m_\text{H}}\rho_{c,0}^{2/3}(1+z)^2 = 
\]
\begin{equation}\label{e10}    
    \simeq 20\left(\frac{1+z}{10}\right)^2\left(\frac{M_H}{10^{10} M_\odot}\right)^{1/3}\left(\frac{\sigma}{10^{-20}\text{\,cm}^2}\right)
    \mbox{\ .}
\end{equation}

\begin{figure}
    \centering
    \includegraphics[width=0.5\textwidth]{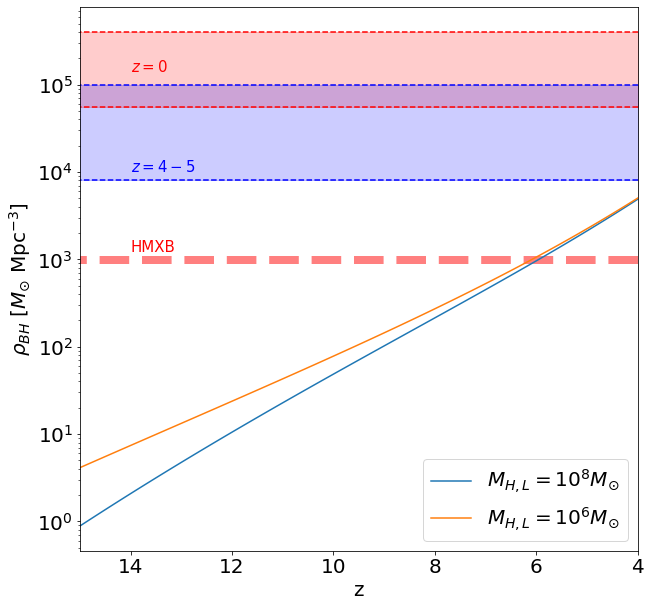}
    \caption{BH mass density vs. redshift for two lower limits for the halo mass function. Assuming that every halo produces a black hole, the integrated halo mass density is converted to a SMBH mass density using Equ.~\ref{e1}, \ref{e4}, and \ref{e9}. The red region indicates the approximate SMBH mass density at $z=0$ \citep{Graham2007}, and the blue region the modeled SMBH density at $z=4-5$ \citep{Mahmood2005,Tucci2017}. For comparison, we also show the extrapolated BH mass density (red dashed line) from accretion onto HMXBs \citep{Jeon2014}. The HMXB contribution here appears as a constant, because \citet{Jeon2014} considered mainly Pop~III binaries at $z\gtrsim 15$, and not the complete populations of metal-enriched stars at lower redshifts. Our model is consistent with observations as the mass density does not exceed the model values at $z\sim4$. } 
    \label{fig1}
\end{figure}

For the cross section, $\sigma$, we consider both Thomson scattering in a fully ionized phase, with a cross section of $\sigma_\text{T}=0.67\times 10^{-24}$\,cm$^2$, and photo-ionization in (substantially) neutral hydrogen gas with \citep{Barkana2001}:
\begin{equation}\label{e11}
    \sigma_\text{p} = 6\times10^{-17}\left(\frac{h_p\nu}{13.6~\text{eV} }\right)^{-3}\text{cm}^2\mbox{\ ,}
\end{equation}
where $h_p$ is the Planck constant and $\nu$ the frequency.
We average $\sigma_\text{p}$ over frequency, weighted by the non-thermal flux given in equ.~5 of \citet{Jeon2014}, and normalizing to the Eddington luminosity. The non-thermal flux refers to the power-law spectral component above a critical frequency, $\nu_\text{NT}$. With frequency limits of $\nu_\text{NT}=0.2$ keV/$h_p$ and 10~keV/$h_p$ \citep[][]{Jeon2014}, the effective photo-ionization cross section is $\sigma_\text{p}\simeq 1.6\times10^{-21}$\,cm$^2$. We consider two limiting cases for the optical depth, corresponding to photon transport in predominantly neutral and fully ionized gas, where the photo-ionization cross section will apply to the neutral part of the halo and the Thomson cross section to the ionized part. 

To account for this two-phase configuration, we evaluate a weighted sum of the photon retention probability, using the two different optical depths:
\begin{equation}\label{optical}
    \langle1-e^{-\tau}\rangle= (1-a)(1-e^{-\tau_\text{T}}) + a(1-e^{-\tau_\text{p}})\mbox{\ .}
\end{equation}
Here, $a$ is the fraction of lines of sight, originating at the center, that intersect the neutral phase, assumed to occupy a torus-like geometry around the central black hole, and $\tau_\text{T,p}$ are the Thomson and photo-ionization optical depths, respectively. This geometry is thought to be established by the vertical outflow from the AGN, with a torus opening angle that varies according to type of the AGN, as well as its luminosity and emission line structure \citep{Kawakatu2011,Marin2012,Sazonov2015,Ichikawa2015}. Such wide range of ionization-cone opening angles is supported by observations as well \citep{He2018}. We specifically assume an average half-opening angle for the torus of 30$^{\circ}$, following the low-luminosity AGN opening angles of \citet{Sazonov2015}, given that low-luminosity sources dominate the overall AGN number density. In Equ.~\ref{optical}, we thus set $a\simeq 1/3$.

From Fig.~\ref{fig1}, it is evident that the resulting SMBH mass density does not overproduce the approximate value estimated for $z=4\sim5$ \citep[e.g.][]{Mahmood2005,Tucci2017}, and is clearly less than the local density at $z=0$ \citep[e.g.][]{Graham2007}.  This is also a consistency check on our simplifying assumption that every halo hosts a SMBH. Even though this may not be strictly correct for increasingly high redshifts, every halo eventually will host a SMBH at later times. Changing our assumption on the SMBH occupation fraction would thus only affect the high-$z$ parts of Fig.~\ref{fig1}, while the lower redshift prediction is quite robust. Furthermore, the choice of lower limit for the halo mass does not make a significant difference at lower redshifts, where typical SMBH host masses are much larger, and where most of our constraints are tested. 

\begin{figure}
    \centering
    \includegraphics[width=0.5\textwidth]{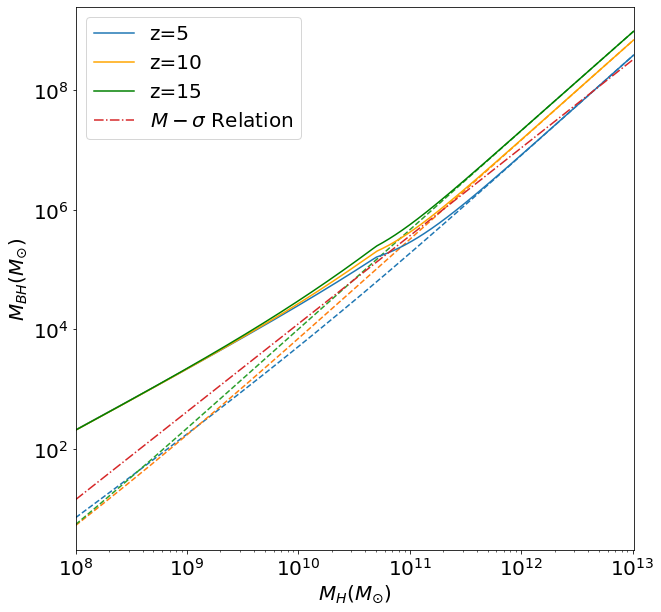}
    \caption{The resulting halo to SMBH mass relation, based on our model (see Equ.~\ref{e1}), and the empirical $M-\sigma$ relation at $z=0$ for comparison \citep{Evrard2008, Kormendy2013}. The dashed lines represent the unmodified Equ.~\ref{e1}, whereas the solid lines show the augmented model where BH seeds are added to the mass budget. As can be seen, our idealized model matches the slope and overall normalization of the observed relation at $z=0$ quite well.} 
    \label{mrelation}
\end{figure}

Fig. \ref{mrelation} shows the relation between host halo mass and SMBH mass within our model, based on Equ.~\ref{e1}, and for comparison also the empirical $M-\sigma$ relation from \citet{Evrard2008} and \citet{Kormendy2013}. Our idealized model is comparable in slope and scale to the observed $M-\sigma$ relation. To avoid SMBH mass values that are unrealistically low, we have also added seed black hole masses to our relation, such that $M_\text{BH}(t)=M_\text{seed}+\dot{M}_\text{acc}t$. The seed masses are set as $M_H\times2\times10^{-6}$, and have lower and upper limits of $10^2$ and $10^5$ M$_\odot$ to not have unrealistic seed black hole masses \citep[e.g.][]{Inayoshi2020}. The effects of the seed black holes are further discussed in Section~\ref{seed} below.

\subsubsection{Stellar X-ray Binaries}
Another important source of X-ray feedback in the early Universe is provided by stellar-remnant X-ray binaries \citep[e.g.][]{Mirabel2011}, in particular high-mass X-ray binaries (HMXBs). A number of studies have focused on the binary statistics of Pop~III stars \citep{Stacy2013, Liu2021}, also extrapolating to the final remnant stage, with considerable uncertainties remaining. As an illustrative example, \citet{Jeon2014} selected around 30\% of Pop~III binaries in their simulation to form HMXBs, following \citet{Power2009}. While \citet{Jeon2014} capture HMXBs from Pop~III, their modeling of subsequent, metal-enriched populations is highly incomplete. Comprehensively determining the contribution from HMXBs to the aggregate mass and energy densities considered here, would require a separate, dedicated study, which is beyond the scope of this paper. With this caveat in mind, we reproduce their HMXB mass accretion density, as shown in Fig.~\ref{fig1}, as a ballpark comparison for our black hole model. We do not however explicitly include the HMXB contribution in our modelling, but only consider SMBH emission.

\begin{figure}
    \centering
    \includegraphics[width=0.5\textwidth]{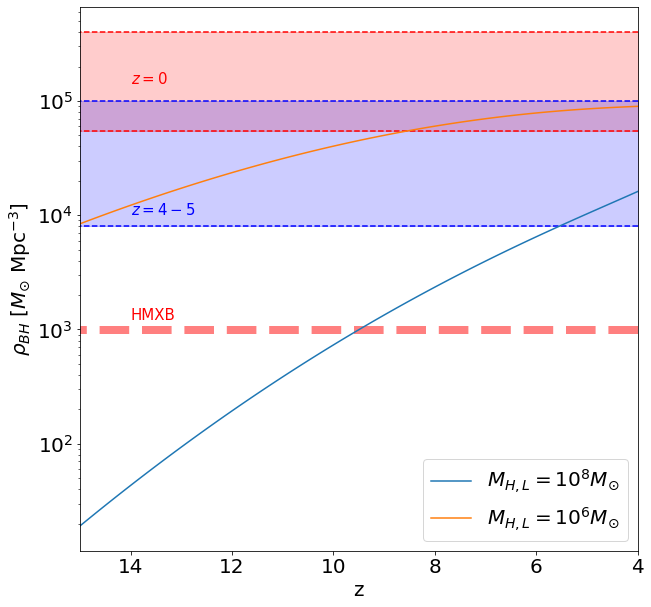}
    \caption{Impact of seed black holes on global mass densities. We consider the same quantities as in Fig.~\ref{fig1}, except that here seed black hole masses are included. Now, the choice of lower limit on the halo mass is reflected in significant differences in the resulting mass densities. Specifically, a lower limit of 10$^6$ M$_\odot$, coupled with a seed BH mass of $\sim 100$\,M$_{\odot}$, overproduces the empirical mass density at $z\simeq 4-5$, and even challenges the present-day constraint. The larger lower limit for the halo mass, on the other hand, still leads to overall consistent results.} 
    \label{seedbh}
\end{figure}

\subsubsection{Seed Black Holes}\label{seed}
We now consider the impact of explicitly modelling seed black holes in the resulting global BH mass densities in somewhat greater detail. Specifically including the mass-dependent seed masses used in producing Fig.~\ref{mrelation}, we show the modified densities in Fig.~\ref{seedbh}, which is to be compared with Fig.~\ref{fig1}, where seed BHs had been ignored. Now, to not violate the density constraints, the lower halo mass limit becomes important. Evidently, for a lower limit of 10$^6$ M$_\odot$, the global mass density in SMBH is overproduced, even challenging the $z=0$ constraint. The lower limit of 10$^8$ M$_\odot$, on the other hand, corresponding to an atomic cooling halo, produces a more realistic mass density. Therefore, the 10$^6$ M$_\odot$ (minihalo) mass limit can be considered as the minimum halo mass required to trigger Pop~III star formation \citep[e.g.][]{Bromm2013}, whereas the minimum halo mass to host a central SMBH is 10$^8$ M$_\odot$, which is of the same order as the suggested minimum mass to host a bona-fide first galaxy \citep[e.g.][]{Bromm2011}.

Furthermore, we conclude that DCBH seeds of $\sim$10$^5$ M$_\odot$ need to be rare, at least compared to our optimistic assumption that every halo hosts a central SMBH, to not overproduce the resulting mass density \citep[also see][and references therein]{ASmith2019}. If we included a $10^5$ M$_\odot$ DCBH seed black hole in every atomic cooling halo, the total BH mass density would be overproduced. More specifically, the fraction of atomic cooling haloes hosting DCBH seeds may be as low as $10^{-2}\sim10^{-7}$ (assuming a comoving number density of atomic cooling haloes of $\sim$10 Mpc$^{-3}$ at $z=10$), depending on the strength of the Lyman-Werner background at that time \citep{Habouzit2016}.  Finally, not all minihaloes near $\sim10^6$ M$_\odot$ can host 100\,M$_\odot$ seeds, as the global mass density would be overproduced in that case, as shown in Fig.~\ref{seedbh}. However, the remaining constraints and tests are based on the radiation output \citep{Soltan1982}, and will thus only depend on the mass accreted onto the SMBH, shown in Fig.~\ref{fig1}. Therefore, the halo lower mass limit will not greatly affect the other tests applied to our model, and choosing 10$^8$ M$_\odot$ as the lower limit is reasonable, as this choice also satisfies the global BH mass density constraint.

\subsection{Energy Density}
To convert the BH mass densities in Fig.~\ref{fig1} to energy densities, we need to calculate the total energy output of an AGN source over its lifetime. The AGN spectral energy distribution is often modeled with two components: a non-thermal (NT) part and a thermal multi-colour disk (MCD) one \citep[e.g.][]{Jeon2012, Jeon2014}. The NT component dominates above $\nu_\text{NT}=0.2 \text{keV}/h_p$, corresponding to the X-ray band. Conversely, the MCD component is responsible for producing the UV energy density. Assuming that the total AGN luminosity is given by the Eddington luminosity, equally shared by the NT and MCD components, the total X-ray and UV luminosities both equal 0.5$L_\text{Edd}$. Further assuming that an AGN source is active for a Salpeter time, the total X-ray/UV energy produced per solar mass in BHs is  
\begin{equation}\label{mtoe}
    E_{\text{X-ray, UV}} \simeq 9\times10^{52}\text{\,erg}\left(\frac{M_{\text{BH}}}{M_{\odot}}\right)\left(\frac{t_\text{Sal}}{5\times10^7 \text{yr}}\right)\mbox{\ .}
\end{equation}
To calculate the energy from X-ray binaries, we follow \citet{Jeon2014}, who also assume the Eddington value for the HMXB bolometric luminosity. Again, half of the luminosity will be in the X-ray band, so that the same equation as above can be used for HMXBs. Their contribution is not included in our figures however as we consider only SMBH radiation.


\section{Cosmological Constraints}\label{3}

\subsection{Reionization}
\begin{figure}
    \centering
    \includegraphics[width=0.5\textwidth]{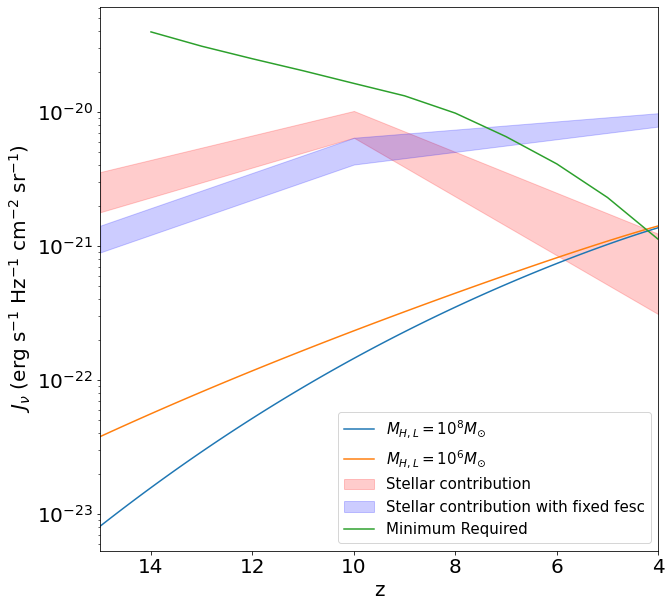}
    \caption{Ionizing radiation intensity vs. redshift. We show the UV intensity produced by AGN sources prior to reionization ({\it blue and orange lines}), together with the intensity from star forming galaxies ({\it shaded blue and red zones}), based on an extrapolation of observations of non-ionizing UV emission from galaxies at $z\simeq 4-10$  \citep{Finkelstein2019}. For the AGN UV contribution, we consider lower limits for the halo mass of 10$^6$ M$_\odot$ (blue line) and 10$^8$ M$_\odot$ (orange line). The stellar contribution is subject to the uncertainty regarding the escape fraction of ionizing photons. The blue shaded region shows the results when a fixed ionizing photon escape fraction of 13\% is used for all galaxies, while the red shaded region shows the results when assuming a halo-mass dependence for the escape fraction. We also indicate the critical intensity required to complete reionization from \citet{Madau1999}, representing the instantaneous emissivity required to sustain a reionized IGM, which depends on the IGM clumping factor ({\it green line}). We assume a clumping factor similar to \citet{Finkelstein2019}. The studies referenced above report UV emissivities, which we convert to photon densities by multiplying with the Hubble time.}
    \label{reionization}
\end{figure}

To test if our model is consistent with the observational constraint that reionization is complete by $z\sim 5.3-5.5$ \citep[e.g.][]{Kulkarni2019,Becker2015}, we compare the UV intensity derived from our SMBH mass density with the minimum intensity required to reionize the Universe \citep{Madau1999}. Specifically, we derive the ionizing intensity from our model as follows:
\begin{equation}
    J_{\text{UV,}\nu}(z)\simeq \frac{ch_p}{4\pi}\frac{u_\text{UV}(z)}{\text{13.6\,eV}}
    \mbox{\ ,}
    \label{JUV}
\end{equation}
where $u_\text{UV}(z)\simeq E_\text{UV}\rho_\text{BH}(1+z)^3$ is the physical energy density in the UV band produced by AGN sources.
In Fig.~\ref{reionization}, we further reproduce the observed intensity from star-forming galaxies from \citet{Finkelstein2019}. Our model does not exceed the minimum required intensity before $z=5-6$, and does therefore not predict an unrealistically early epoch of reionization, which would be in conflict with observations. Our model also predicts that X-ray and stellar sources made similar contributions to the ionizing photon budget by $z\sim5$, not long after the end of reionization. 

\subsection{X-ray Luminosity Function}
\begin{figure}
    \centering
    \includegraphics[width=0.5\textwidth]{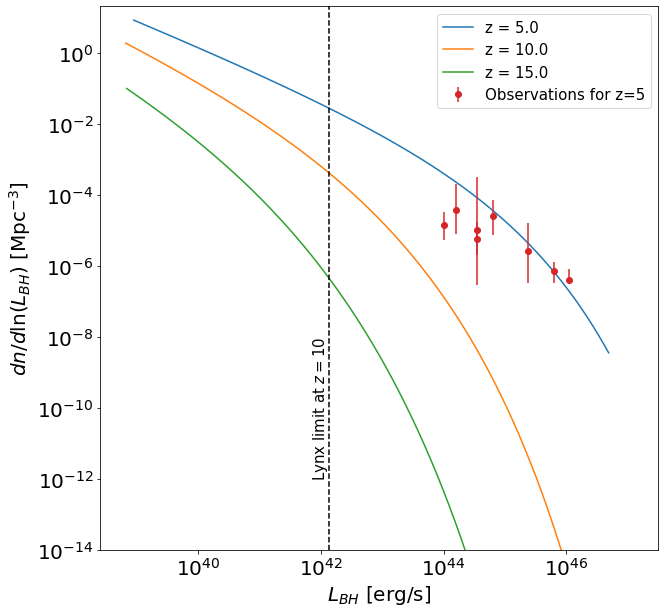}
    \caption{AGN luminosity function at high redshift. We show comoving number densities versus bolometric luminosity, derived from our PS mass functions for select redshifts. The observed luminosity function at $z=5$ is reproduced from \citet{Shen2020}. We also mark the limiting luminosity that can be reached with the Lynx observatory at $z=10$ \citep{Ricarte2018}. As can be seen, our luminosity function at $z=5$ is in overall agreement with the observations.}
    \label{luminosity}
\end{figure}

We convert our PS halo mass functions, $dn/d\ln M_H$, to luminosity functions at different redshifts, expressed in comoving units and assuming that sources emit at the Eddington luminosity: 
\begin{equation}
    \frac{dn}{d\ln L_\text{BH}}\simeq \frac{dn}{d\ln M_\text{BH}}
    \simeq \frac{3}{5} \frac{dn}{d\ln M_H}\mbox{\ .}
\end{equation}
Note that the factor in the second equality here arises from the halo mass dependence in Equ.~\ref{e4}. In
Fig.~\ref{luminosity}, we compare our derived bolometric luminosity functions to observations at $z=5$ \citep{Shen2020}. The predicted function at $z=5$ is broadly consistent with the observations. It is currently not possible to directly probe the AGN luminosity function at the highest redshifts, leaving only the high-luminosity tail at the lowest redshifts modeled here accessible to observations.\footnote{However, a number of additional empirical constraints on the quasar luminosity function at $z\sim6$ exist \citep[e.g.][]{Jiang2016}.} Future facilities, such as Lynx, Athena, and AXIS, will push to significantly fainter limits, promising a much more detailed scrutiny of the high-$z$ AGN population. In particular, these frontier facilities will be able to observe X-ray luminosities as low as $10^{42}\sim10^{43}$ erg s$^{-1}$ near $z\sim6$ \citep{Marchesi2020,Habouzit2022}.

\subsection{Soft X-ray background}
\begin{figure}
    \centering
    \includegraphics[width=0.5\textwidth]{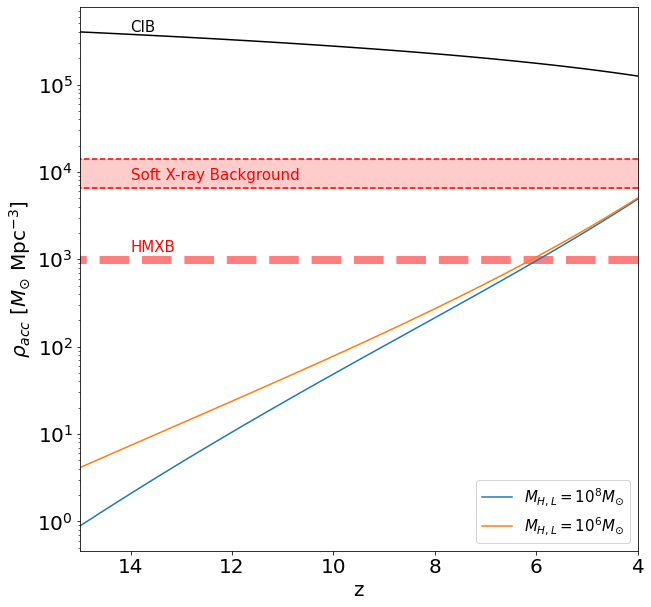}
    \caption{Cosmic radiation backgrounds from SMBH mass accretion. Similar to Fig.~\ref{fig1}, we show our model SMBH mass accretion densities, compared to upper limits from cosmological backgrounds. The soft X-ray background limit corresponds to the maximal possible accretion for $z \gtrsim 5$, based on the unresolved background from \citet{Salvaterra2012}. The CIB limit is the mass accretion required for SMBHs to produce the observed CIB fluctuations at a given redshift, provided in \citet{Helgason2016}. }
    \label{soft}
\end{figure}

The soft X-ray background is the lower energy part of the X-ray background ($\leq2$ keV) with contributions mainly from AGNs, but also from galaxies and stellar sources \citep{Mcquinn2012,Cappelluti2012}. While most of the background has been resolved into point and extended sources with \textit{Chandra} and \textit{XMM–Newton} \citep{Moretti2003,Bauer2004,Lehmer2012}, a sizeable 5--10\% of the background remains unresolved. Low-luminosity AGNs most likely make up the bulk of this unresolved fraction, and for consistency, our model cannot overproduce the energy deposited into the X-ray background. For simplicity, we assume that the entire BH mass density, as shown in Fig.~\ref{fig1}, has been assembled through radiatively efficient accretion, such that $\rho_\text{acc}\sim \rho_\text{BH}$. We thus effectively neglect any contribution to the BH mass budget from seed BHs and mergers. At least the former approximation is well justified, as discussed above. With these idealized assumptions, we plot SMBH mass accretion densities in Fig.~\ref{soft}, and compare to the upper limit for the unresolved soft X-ray background, expressed in the same units \citep{Salvaterra2012}. As can be seen, our model does not exceed this limit even at $z=4$, demonstrating overall consistency with the X-ray background observations.

\subsection{Cosmic infrared background}
X-ray feedback contributes indirectly to the cosmic infrared background, as well. More specifically, recombining gas that is ionized by X-rays produces Ly$\alpha$ photons. The Ly$\alpha$ radiation, emitted at $z\sim 10$ is then redshifted into the infrared by $z=0$, thus contributing to the CIB today \citep{Rybicki1986, Santos2002}. The current limits on the unresolved CIB, therefore, impose another global constraint on pre-reionization X-ray feedback. To illustrate this additional constraint, Fig~\ref{soft} also shows the accretion density required for SMBHs to produce the observed CIB fluctuations, given by equation~16 in \citet{Helgason2016}. Our model accretion lies far below this value, thus obeying the CIB constraint, but also predicting that SMBHs alone cannot account for the observed level of CIB fluctuations.

\subsection{21 cm radiation}
\begin{figure}
    \centering
    \includegraphics[width=0.5\textwidth]{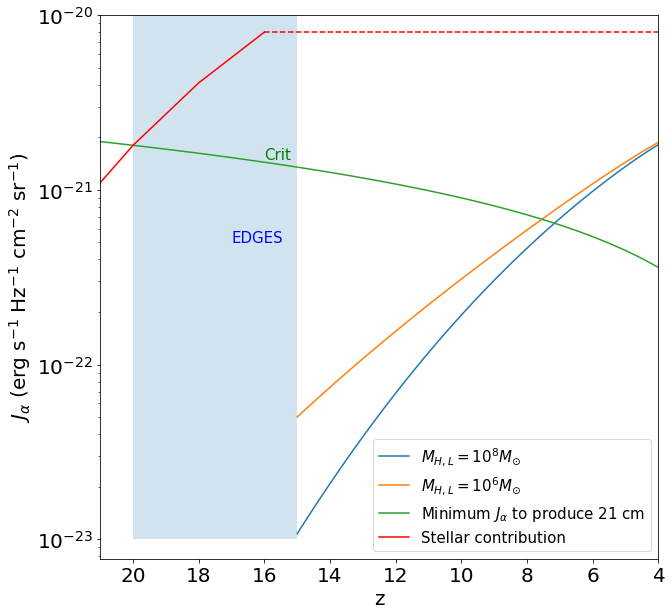}
    \caption{The cosmic Lyman-$\alpha$ intensity vs. redshift, from stellar sources (galaxies) and AGN feedback, similar to Fig.~\ref{reionization}. We reproduce the minimum Ly$\alpha$ intensity required to imprint a global 21\,cm absorption feature, taken from \citet{Ciardi2003_2}. We also show the stellar contribution, mostly from Pop~III, as modeled in \citet{Schauer2019}. The EDGES timing constraint from \citet{Mirocha2019} is indicated with the blue shaded region. The stellar contribution at $z=16$ is extrapolated to $z=4$ (dashed line), to compare to the AGN contribution modeled here. Our model does not reach the necessary intensity until $z\sim 8$, and thus does not produce the 21\,cm line too early.}
    \label{21}
\end{figure}

In general, X-ray sources indirectly contribute to the cosmological 21\,cm signature \citep[e.g.][]{Furlanetto2006}. The X-ray emission heats the surrounding neutral hydrogen gas and produces Ly$\alpha$ photons. The total Ly$\alpha$ radiation, including all available sources, is then able to produce observable 21\,cm lines from the Wouthuysen–Field effect \citep{Field1958,Wouthuysen1952}. Here, neutral hydrogen absorbs Ly$\alpha$ photons and produces a change in spin state during relaxation, thus rendering the 21\,cm transition detectable in absorption relative to the CMB \citep{Pritchard2012}. In Fig.~\ref{21}, we compare the Ly$\alpha$ intensity from our model to the critical intensity required to produce the 21\,cm line via the Wouthuysen-Field coupling \citep{Ciardi2003_2}. For simplicity, we assume that each ionization event results in the emission of a Lyman-$\alpha$ photon, such that $J_\alpha\simeq J_{\text{UV},\nu}$ (see Equ.~\ref{JUV}). A detailed modeling of the complex radiative transfer and nebular reprocessing involved is beyond the scope of this paper\footnote{The standard nebula physics of case-B recombination predicts that only 0.68 of all recombinations would result in the emission of a Ly$\alpha$ photon. Such order-unity corrections would not change any of our conclusions.}. For comparison with the AGN contribution, we also reproduce the Ly$\alpha$ intensity originating in massive Pop~III stars \citep{Schauer2019}. Our model does not reach the minimum intensity until $z=8$, while the stellar intensity does at $z=20$.

The moment of `cosmic dawn', when Wouthuysen-Field coupling first becomes effective, is currently under vigorous debate. The Experiment to Detect the Global Epoch of Reionization Signature (EDGES) had reported a strong, global absorption feature at 78\,MHz, corresponding to a redshift of $z\sim 17$ for the line centre \citep{Bowman2018}. We indicate the EDGES constraint in Fig.~\ref{21}, as well, implying that our modeled AGN intensity does not produce the 21\,cm line too early, and is thus consistent with observations. We note that the recent result from the Shaped Antenna Measurement of the Background Radio Spectrum (SARAS) has cast doubt on the original EDGES detection \citep{Singh2022}, but the radio-astronomical debate is far from over.

Besides the Wouthuysen-Field effect, X-ray feedback can affect the 21\,cm signal through pre-heating of the diffuse gas. Specifically, HMXBs have been shown to pre-heat the IGM and increase the brightness temperature on large scales \citep[e.g.][]{Jeon2014}. The overall effect is that the heated IGM, outside the regions that are ionized by stellar sources, can produce the 21\,cm line at early times, whereas the fully-ionized IGM at later times cannot \citep{Baek2010}. However, this effect is difficult to assess with our current model considering the complex physics of the IGM and the clustering of AGN sources. A cosmological simulation will be able to study this heating effect comprehensively.

\section{Summary and Conclusions}\label{5}

In this paper, we have constructed a model to explore the impact of X-ray feedback from high-redshift AGN on the Universe before reionization. X-ray feedback can have conflicting effects on reionization and early star formation, so we assess limits on the available X-ray feedback within our model. We follow a semi-analytical approach, starting with the Press-Schechter halo mass function \citep{Press1974}, and imposing a balance between the dark matter halo gravitational potential energy and the feedback energy from a central SMBH, to derive a relationship between the halo and black hole mass. We further use the halo-SMBH mass relation to predict SMBH mass densities over cosmic time. We in particular choose our model parameters to produce the maximum amount of X-ray feedback possible. The SMBH occupation fraction is set to unity, so that every halo above a critical mass ($10^6$ or $10^8$ M$_\odot$) hosts at least one SMBH for maximal feedback. 

We subject our model to a number of tests to see if it violates any observational constraints. Among the tests, we consider the global SMBH mass density, the completion of reionization, the quasar luminosity function, select cosmic radiation backgrounds, and the (redshifted) 21\,cm line, imprinted by neutral hydrogen in the early IGM. Regarding the mass density, luminosity function, and the cosmic backgrounds considered, our model does not overproduce the current observational constraints. Regarding reionization and the 21\,cm line, our model does not produce the corresponding couplings between radiation fields and hydrogen gas too early, and is thus also consistent with these observations. Our X-ray feedback model, thus tested and constrained, can serve as a sub-grid model in future cosmological simulations of high-$z$ galaxy formation.

In our derivation of the model, only a few free parameters remain including the SMBH occupation fraction and the 3D geometry of \HII\ regions in the halo. However, modifying the free parameters may contradict the empirical constraints. Our model predictions are close to the observed upper limits on the luminosity function and the soft X-ray background, as shown in Figures~\ref{luminosity} and \ref{soft}, such that the feedback cannot be much higher. We further argue that our choice for these parameters have physical merit. The chosen ionized fraction of the halo is in line with previous work on AGN opening angles \citep{Sazonov2015}. The occupation fraction is most likely unity at lower redshifts, as every observed galaxy hosts a black hole at its center \citep{Kormendy2013}. Although deviations from our simple `one SMBH per halo' assumption are likely at high redshifts, imposing this condition on all haloes and cosmic times is adequate for our current semi-analytic model.
 
We seemingly have little freedom in designing our subgrid model, because the model parameters are tightly constrained by seemingly robust physical arguments. Within the original energy feedback argument, the dimensionless parameter $f$ is effectively set by assuming Eddington-limited accretion, occurring over the Salpeter timescale. However, some previous studies have argued that in the early Universe, SMBHs went through super-Eddington accretion instead \citep[e.g.][]{Pezzulli2016,Regan2019}. We have also assumed a simple, bipolar geometry for the ionized \HII\ regions, whereas the true situation is likely much more complex, requiring sophisticated 3D radiative transfer simulations to fully capture reality \citep{Alvarez2007,McQuinn2007}. Nevertheless, our idealized model serves to explore the relevant parameter space, thus informing follow-up cosmological simulations.  

Our modeling is subject to a number of caveats. Clearly, our treatment of SMBH accretion and the structure of the ionized halo regions is highly idealized. The constraints mostly derive from observations at lower redshifts, so that the high-$z$ modelling is only indirectly tested. For example, the most direct constraints considered here, the SMBH mass density and the quasar luminosity function, are currently observed only at low redshifts \citep{Shen2020}, and even the constraint at $z=4-5$ on the BH mass density involves model extrapolations \citep{Mahmood2005,Tucci2017}. Even the most robust tests therefore exhibit considerable uncertainties, which will be addressed with future observations at higher redshifts. 

Overall, our model indicates that X-ray feedback was important in the evolution of the early Universe. While the model X-ray feedback does not significantly contribute to the CIB or the 21\,cm line, Fig. \ref{reionization} indicates that the contribution from AGN may have been comparable to stellar sources in completing the reionization of the Universe. Furthermore, Fig. \ref{soft} implies that high-$z$ AGNs could explain the unresolved soft X-ray background. Thus, excluding X-ray feedback when studying early cosmic evolution will not be realistic. In contrast, previous work such as \citet{Cappelluti2017} concluded that the emission from high-$z$ AGN sources would have to be heavily obscured in Compton-thick envelopes to not violate existing constraints. In this case, the large optical depth would significantly reduce the impact of any AGN feedback. In future work, employing cosmological simulations, we will more thoroughly explore such radiative transfer effects. To improve our modelling of X-ray feedback, direct observations of high-redshift AGN/SMBH sources will be needed. More specifically, distinguishing the feedback from AGN and stars at high redshifts may be possible with future X-ray observatories, such as Athena and Lynx \citep{Barret2013,Ricarte2018,Gaskin2019}, or with observations of the rest-frame UV at different wavebands with \textit{JWST} \citep{Volonteri2017}. 
\section*{Data availability}
The data underlying this article will be shared on reasonable request to the corresponding author.





\bibliographystyle{mnras}
\bibliography{ms} 





\bsp	
\label{lastpage}
\end{document}